\documentclass[12pt]{article}
\usepackage{putex}
\usepackage{graphicx}
\usepackage{epstopdf}
\usepackage{enumerate}
\usepackage{cite}

\newcommand{\abs}[1]{\lvert #1 \rvert}

\newcommand {\be} {\begin {equation}}
\newcommand {\ee} {\end {equation}}

\newcommand {\bes} {\begin {equation*}}
\newcommand {\ees} {\end {equation*}}

\newcommand{\es}[2] {\begin{equation} \label{#1} \begin{split} #2 \end{split} \end{equation}}

\newcommand{\Z}{\mathbb{Z}}
\newcommand{\R}{\mathbb{R}}
\newcommand{\C}{\mathbb{C}}

\begin{document}

\preprint{PUPT-2343}

\institution{PU}{Joseph Henry Laboratories$^1$ and Center for Theoretical Science,$^2$\cr
~~~~~~~~~~~~~~Princeton University, Princeton, NJ 08544}

\title{M-Branes and Metastable States}

\authors{Igor R. Klebanov$^{1,2}$ and Silviu S. Pufu$^1$}

\abstract{
We study a supersymmetry breaking deformation of the M-theory background found in arXiv:hep-th/0012011.
The supersymmetric solution is a warped product of $\R^{2,1}$ and the 8-dimensional Stenzel space, which is a higher dimensional generalization of the deformed conifold. At the bottom of the warped throat there is a 4-sphere threaded by $\tilde M$ units of 4-form flux.  The dual $(2+1)$-dimensional theory has a discrete spectrum of bound states. We add $p$
anti-M2 branes at a point on the 4-sphere, and show that they blow up into an M5-brane wrapping a 3-sphere at a fixed azimuthal angle on the 4-sphere. This supersymmetry breaking state turns out to be metastable for $p/\tilde M \lesssim 0.054$. We find a smooth $O(3)$-symmetric Euclidean bounce solution in the M5-brane world volume theory that describes the decay of the false vacuum. Calculation of the Euclidean action shows that the metastable state is extremely long-lived. We also describe the corresponding metastable states and their decay in the type IIA background obtained by reduction along one of the spatial directions of $\R^{2,1}$. 
}

\date{June 2010}

\maketitle

\tableofcontents

\newcommand{\const}{\lambda}

\section{Introduction}

An important general question about supersymmetric field theories is whether they admit long-lived metastable states that break supersymmetry. Constructions of such states often require that the supersymmetric field theory is strongly coupled. In these cases non-perturbative phenomena, such as the Seiberg duality \cite{Seiberg:1994pq}, can sometimes be used to demonstrate the possibility of metastable supersymmetry breaking \cite{Intriligator:2006dd}.

Another tool available for studying strongly coupled gauge theories is the AdS/CFT duality \cite{Maldacena:1997re,Gubser:1998bc,Witten:1998qj}. In this context, the background dual to a metastable state should be a locally stable non-supersymmetric solution which is asymptotic to a supersymmetric AdS background of string theory or M-theory. The first construction of a string dual of a metastable state was presented by Kachru, Pearson, and Verlinde (KPV) \cite{Kachru:2002gs} in the context of the warped deformed conifold background \cite{Klebanov:2000hb} of type IIB string theory. In the ultraviolet this background is close to $AdS_5 \times T^{1,1}$ up to logarithmic effects that encode the running of the couplings and the cascade of Seiberg dualities \cite{Klebanov:2000nc}. In the infrared the throat ends smoothly with the warp factor approaching a finite value which signals the color confinement. The internal space contains a blown-up 3-sphere threaded by $M$ units of R-R 3-form flux. In the absence of additional space-time filling branes, the infrared ${\cal N}=1$ supersymmetric theory has gauge group $SU(M)\times SU(2M)$ coupled to bi-fundamental chiral superfields. The KPV construction involves adding $p$ coincident anti-D3 branes that break supersymmetry and are attracted to the bottom of the throat.\footnote{A further important problem, which was addressed in \cite{DeWolfe:2008zy,Bena:2009xk}, is finding the back-reaction of the anti-D3 branes.}
There, due to the R-R 3-form flux, they get blown up via the Myers effect \cite{Myers:1999ps,Bachas:2000ik} into an NS5-brane wrapping a 2-sphere located at a fixed azimuthal angle within the 3-sphere; the NS5-brane carries $-p$ units of world volume flux that endows it with the D3-brane charge. An explicit calculation \cite{Kachru:2002gs} of the potential as a function of the angle shows that for $p/M \lesssim 0.08$ this state is metastable with respect to decay via ``brane-flux annihilation'' to the supersymmetric ground state of the $SU(M-p)\times SU(2M-p)$ gauge theory. Estimates of the decay rate via tunneling show that this state is typically extremely long-lived \cite{Kachru:2002gs,Freivogel:2008wm}.

In this paper we present an analogous construction of long-lived metastable states in the ${\cal N}=2$ supersymmetric $(2+1)$-dimensional theory dual to the $AdS_4\times V_{5,2}$ background of M-theory supported by $N$ units of 4-form flux. The 7-dimensional Sasaki-Einstein space $V_{5,2}=SO(5)/SO(3)$ is the base of the conical CY 4-fold $\sum_{i = 1}^5 {z_i^2} = 0$ \cite{Klebanov:1998hh}. The Kaluza-Klein spectrum of the $AdS_4\times V_{5,2}$ background, and a proposal for the dual gauge theory, were originally discussed in \cite{Ceresole:1999zg}. Recently, two different new proposals for the dual gauge theory were made \cite{Martelli:2009ga,Jafferis:2009th}. The first of them \cite{Martelli:2009ga} is an ${\cal N}=2$ supersymmetric $U(N)\times U(N)$ Chern-Simons gauge theory, quite analogous to the ABJM theory \cite{Aharony:2008ug}. This gauge theory is strongly coupled because it involves the minimal Chern-Simons levels $(1,-1)$.  A rather different strongly coupled gauge theory, involving a $U(N)$ gauge group coupled to adjoint and fundamental matter, was suggested in \cite{Jafferis:2009th}.
On the other hand, at large $N$, the dual M-theory description is weakly coupled. This allows us to search for metastable states using quasi-classical methods. Another crucial fact is that there exists a natural deformation of the gauge theory whose weakly curved M-theory dual was found by Cvetic, Gibbons, Lu, and Pope (CGLP) \cite{Cvetic:2000db}; it is a warped product of $\R^{2, 1}$ and the eight-dimensional Stenzel space \cite{Stenzel} $\sum_{i = 1}^5 {z_i^2} = \epsilon^2$, which is a higher-dimensional generalization of the deformed conifold \cite{Candelas:1989js}. The CGLP background \cite{Cvetic:2000db} is similar to the KS solution \cite{Klebanov:2000hb}, except it is asymptotic to $AdS_4\times V_{5,2}$ without any UV logarithms. In the infrared the background contains a blown-up 4-sphere threaded by
$\tilde M$ units of 4-form flux (as shown in \cite{Martelli:2009ga}, $N=\tilde M^2/4$), and the warp factor approaches a finite value.
Thus, the background has a discrete spectrum of normal modes which describe bound states in the dual field theory \cite{JLThesis, Pufu:2010ie}.
Some aspects of the infrared physics were discussed in \cite{Herzog:2000rz,Martelli:2009ga}, but the dual infrared gauge theory remains to be elucidated.

Our M-theory construction of metastable states involves adding $p$ anti-M2 branes that fall to the bottom of the CGLP warped throat (for the proposal of \cite{Martelli:2009ga}, the UV conformal gauge theory is then $U(N-p)\times U(N-p)$). The 4-form flux blows the anti-M2 branes up into a single M5-brane wrapping a 3-sphere located at a fixed azimuthal angle inside the 4-sphere. Our explicit calculation of the potential as a function of the angle shows that this state is metastable for
$p/\tilde M \lesssim 0.054$. We construct a smooth $O(3)$-symmetric Euclidean bounce solution for the M5-brane world volume theory that describes the decay of the false vacuum \cite{Coleman:1977py}: the inner region is near the true vacuum while the outer region is in the false vacuum. Calculation of the Euclidean action shows that the metastable state is extremely long-lived. Nevertheless, for $p$ not too small our result deviates significantly from the thin-wall limit that has been used in the literature  \cite{Kachru:2002gs,Freivogel:2008wm}.
In the present case, the thin domain wall is the M5-brane wrapped over the 4-sphere. Due to the $\tilde M$ units of 4-form flux through the $S^4$, there must be $\tilde M$ M2-branes ending on the domain wall. Thus, the domain wall interpolates between the non-supersymmetric state containing $p$ anti-M2 branes and a supersymmetric state with $\tilde M-p$ M2-branes. For $p=0$ this domain wall becomes BPS and interpolates between two supersymmetric vacua.

We also discuss an analogous type IIA construction where the CGLP solution is compactified along one of the longitudinal directions,
$x^2$.  In this case, the metastable state corresponds to adding $p$ fundamental strings oriented in such a way that they break supersymmetry; they blow up into a D4-brane wrapping a 3-sphere inside the 4-sphere. In this case, the metastable vacuum decays via nucleation of a D4-brane and an anti-D4 brane wrapping the 4-sphere. We treat the tunneling amplitude for this Schwinger process relativistically and show that the result agrees with the Euclidean approach where we obtain the requisite $O(2)$-symmetric solution.

The structure of the paper is as follows. In section 2 we review the CGLP solution \cite{Cvetic:2000db} of 11-dimensional supergravity and also present some new results. In particular, we express the forms $\sigma_i, \tilde \sigma_i$, and $\nu$ in terms of the 7 angular coordinates of $V_{5,2}$ found in \cite{Bergman:2001qi}; this allows us to obtain an explicit form of the Stenzel metric. We also put the 4-form field strength in a manifestly $SO(5)$-invariant form. In section 3 we present a reduction of the CGLP background to type IIA along one of the
spatial coordinates of $\R^{2, 1}$. In section 4 we study the UV ($\tau\rightarrow \infty$) and IR ($\tau\rightarrow 0$) limits of the backgrounds. We also discuss the BPS domain walls made of D4 or M5 branes wrapping the $S^4$ at $\tau=0$. In section 5 we show that strings placed at the bottom of the type IIA background can blow up into a D4-brane wrapping an $S^3$ inside the $S^4$; we also carry out the analogous calculation for anti-M2 branes at the bottom of the M-theory background. In section 6 we calculate the semiclassical decay rate of these metastable states. In section 7 we conclude and discuss some open problems. In Appendix A we show that the blown-up brane does not reside at $\tau>0$.

\section{Review of the CGLP background}
\label{REVIEW}

\subsection{The eight-dimensional Stenzel space}

The eleven-dimensional supergravity background \cite{Cvetic:2000db} is given by a warped product of $\R^{2, 1}$ and the eight-dimensional Stenzel space, which is a deformed cone over the Sasaki-Einstein space $V_{5, 2}$.  Let us review the construction of this background.

We start with the (undeformed) cone over the Sasaki-Einstein manifold $V_{5, 2}$, which is a four-complex dimensional space that generalizes the three-complex dimensional conifold \cite{Candelas:1989js}.  It is described by the subset of $\C^5$ given by
\es{UndeformedCone}{
\sum_{i = 1}^5 {y_i^2} = 0 \,,
}
where $y_i$ are the five complex coordinates on $\C^5$.  A Ricci flat metric on this cone which respects the $SO(5)$ symmetry that acts on the $y_i$ can be derived from a K\"ahler potential of the form $\left( \sum_{i = 1}^5 \abs{y_i}^2 \right)^{3\over 4}$.  The base of this cone, namely the Stiefel manifold $V_{5, 2}$, can be found by intersecting \eqref{UndeformedCone} with the unit sphere in $\C^5$,
\es{UnitSphere}{
\sum_{i = 1}^5 \abs{y_i}^2 = 1 \,.
}
Topologically, $V_{5, 2}$ is an $S^3$ bundle over $S^4$, and the cone \eqref{UndeformedCone} is not smooth at the tip where $y_i = 0$.  $V_{5, 2}$ is {\it not} a product space $S^3\times S^4$ because the $S^3$ fiber bundle over $S^4$ is not trivial (see, for example, \cite{Hatcher}).  In this respect $V_{5,2}$ differs from its lower dimensional analog $T^{1, 1} = V_{4, 2}$, which is topologically $S^2\times S^3$.  The only non-trivial cycle in $V_{5, 2}$ is a $\Z_2$ torsion $3$-cycle, which is represented by the $S^3$ fiber.

One can resolve the conical singularity by deforming the cone, i.e.~by replacing it with the Stenzel space \cite{Stenzel}
\es{DeformedV52}{
\sum_{i = 1}^5 z_i^2 = \epsilon^2 \,,
}
where $\epsilon$ is a deformation parameter that can be taken to be real without loss of generality.  A convenient way of parameterizing this deformed space\footnote{We thank J.~Lin and T.~Klose for useful discussions about possible parameterizations of this space.} is by using a ``radial'' coordinate $\tau$ and the coordinates $y_i$ that, subject to the constraints \eqref{UndeformedCone} and~\eqref{UnitSphere}, parameterize $V_{5, 2}$: by writing
\es{zToy}{
z_i = {\epsilon \over \sqrt{2}} \left(e^{\tau \over 2} y_i + e^{-{\tau \over 2}} \bar y_i \right)  \,,
}
it is easy to see that \eqref{DeformedV52} follows from \eqref{UndeformedCone} and~\eqref{UnitSphere}.  In order to cover the deformed conifold only once, the range of $\tau$ should be taken to be from zero to infinity.  At $\tau = 0$, eq.~\eqref{DeformedV52} shows that the space reduces to an $S^4$ of finite size. In fact, the Stenzel space \eqref{DeformedV52} is topologically the tangent bundle\footnote{The tangent bundle $TS^4$ and the cotangent bundle $T^*S^4$ are homeomorphic as $8$-manifolds.} $TS^4$ of $S^4$, where, for example, $\Re y_i$ parameterize the $S^4$ base and $\tau \Im y_i$ parameterize the tangent vectors to it.  Indeed, \eqref{UndeformedCone} and \eqref{UnitSphere} imply that
\es{Candgen}{\sum_{i = 1}^5 (\Re y_i)^2 = \sum_{i = 1}^5 (\Im y_i)^2 = 1/2\,, \qquad
\sum_{i = 1}^5 \Re y_i \Im y_i = 0\,,}
so $\sqrt{2} \Im y_i$ parameterize unit tangent vectors to $S^4$, and $\tau$ is the radial coordinate in the tangent space.  Replacing the $\R^4$ fiber in $TS^4$ by the unit 3-sphere $S^3 \subset \R^4$, one recovers the Stiefel manifold $V_{5, 2}$, so any constant $\tau>0$ section of the Stenzel space is topologically $V_{5, 2}$.

The Stenzel space \eqref{DeformedV52} is a non-compact Calabi-Yau four-fold which we will denote by ${\cal C}_\epsilon$. Its Calabi-Yau metric \cite{Stenzel,Cvetic:2000db} can be derived from a K\"ahler potential $K$ that depends only on $\tau$.  In particular,
\es{MetricDeformed}{
ds_8^2 = K'(\rho) dz_i d\bar z_i + K''(\rho) \abs{z_i d\bar z_i}^2 \,, \qquad
 \rho \equiv \sum_{i = 1}^5 \abs{z_i}^2 = \epsilon^2 \cosh\tau \,,
}
and the K\"ahler potential can be found from integrating the expression
\es{KahlerDeformed}{
K'(\rho) = {\sqrt{2} \const^2 \over 3^{1 \over 4}} {\left(\rho + 2 \epsilon^2 \right)^{1 \over 4} \over \sqrt{\rho + \epsilon^2}} \,,
}
where $\const$ is an arbitrary constant.  The reason why \eqref{KahlerDeformed} contains an arbitrary multiplicative factor $\const^2$ is that any rescaling of a Ricci flat metric is also Ricci flat.

Using the parameterization of $y_i$ from \cite{Bergman:2001qi},\footnote{There is a typo in the expression for $\Lambda_-$ in \cite{Bergman:2001qi}.  The correct expression is $\Lambda_- = \cos \alpha \sin {\beta\over 2} - i \cos {\beta\over 2}$.  We thank C.~Herzog for sending us the corrected formula.} let us write down this Ricci flat metric explicitly.  In \cite{Bergman:2001qi}, the $y_i$ are expressed in terms of seven angles $(\alpha, \beta, \theta_1, \phi_1, \theta_2, \phi_2, \psi)$.  It is convenient to define the quantities
\es{stDefs}{
s_+ &\equiv {1 \over 2} \left[\cos \alpha \cos \psi \sin {\beta\over 2} + \sin \psi \cos {\beta\over 2} \right] \,, \qquad
s_- \equiv {1 \over 2} \left[\cos \alpha \cos \psi \cos {\beta\over 2} - \sin \psi \sin {\beta\over 2} \right] \,,\\
t_+ &\equiv {1 \over 2} \left[\cos \psi \sin {\beta \over 2} + \cos \alpha \sin \psi \cos {\beta \over 2} \right] \,, \qquad
t_- \equiv {1 \over 2} \left[\cos \psi \cos {\beta \over 2} - \cos \alpha \sin \psi \sin {\beta \over 2} \right] \,,
}
and the differential one-forms
\es{DiffForms}{
e_\beta &\equiv d \beta - \cos \theta_1 d\phi_1 - \cos \theta_2 d \phi_2\,, \\
\nu &\equiv -d \psi - {1 \over 2} \cos \alpha\, e_\beta\,, \\
\sigma_1 &\equiv \cos \psi d\alpha + {1 \over 2} \sin \psi \sin \alpha \,  e_\beta \,, \\
\sigma_2 &\equiv -s_+ (d \theta_1 - d \theta_2) - s_- (\sin \theta_1 d\phi_1 - \sin \theta_2 d \phi_2) \, \\
\sigma_3 &\equiv -s_- (d \theta_1 + d \theta_2) + s_+ (\sin \theta_1 d\phi_1 + \sin \theta_2 d\phi_2) \,\\
\tilde \sigma_1 &\equiv \sin \psi d \alpha - {1 \over 2} \cos \psi \sin \alpha \, e_\beta \,, \\
\tilde \sigma_2 &\equiv t_- (d \theta_1 - d\theta_2) - t_+ (\sin \theta_1 d \phi_1 - \sin \theta_2 d\phi_2) \,,\\
\tilde \sigma_3 &\equiv -t_+ (d\theta_1 + d\theta_2) - t_-(\sin \theta_1 d\phi_1 + \sin \theta_2 d\phi_2) \,.
}
The metric on the deformed cone over $V_{5, 2}$ is then
\es{ExplicitMetric}{
ds_8^2 = c^2 \left({1 \over 4} d\tau^2 + \nu^2 \right)
+ b^2 \sum_{i = 1}^3 \tilde \sigma_i^2
+ a^2 \sum_{i = 1}^3 \sigma_i^2 \,,
}
with $a$, $b$, and $c$ being functions only of the radial coordinate $\tau$:
\es{Gotabc}{
a^2 &= \epsilon^2 \cosh^2 {\tau \over 2} K'(\rho)
= {\const^2 \epsilon^{3 \over 2} \over 3^{1 \over 4}} (2 + \cosh \tau)^{1 \over 4} \cosh {\tau\over 2} \,, \\
b^2 &= \epsilon^2 \sinh^2 {\tau \over 2} K'(\rho)
= {\const^2 \epsilon^{3 \over 2} \over 3^{1 \over 4}} (2 + \cosh \tau)^{1\over 4} {\sinh^2 {\tau\over 2}
\over \cosh{\tau\over 2}} \,, \\
c^2 &= \epsilon^2 \cosh \tau K'(\rho) + \epsilon^4 \sinh^2 \tau K''(\rho)
= 3^{3 \over 4} \const^2 \epsilon^{3 \over 2} { \cosh^3 {\tau\over 2}  \over (2 + \cosh \tau)^{{3 \over 4}} }\,.
}
While the form (\ref{ExplicitMetric}) of the metric on the Stenzel space is well-known \cite{Cvetic:2000db}, our new expressions (\ref{DiffForms}) for the angular forms make it perfectly explicit.
Setting $\const = 3^{-{3 \over 8}} \epsilon^{-{3 \over 4}}$ one recovers the conventions of \cite{Cvetic:2000db, Martelli:2009ga}, but this normalization constant should not affect any observable quantities.  We find it convenient to set instead
\es{lambdaChoice}{
\const = \sqrt{3 \over 2} \,.
}

Let us comment on the $\tau \to \infty$ and $\tau = 0$ limits of the metric.  At large $\tau$, the deformed cone ${\cal C}_\epsilon$ should approach the undeformed cone over $V_{5, 2}$.  One can define the standard radial coordinate $r$ of the cone by the relation
\es{rLargetau}{
\rho =  {3^{1 \over 3} \over 4}  r^{8 \over 3} \,.
}
In terms of this coordinate, the metric \eqref{ExplicitMetric} becomes approximately
\es{LargetauMetric}{
ds_8^2 = dr^2 + r^2 \left[{9 \over 16} \nu^2 + {3 \over 8} \sum_{i = 1}^3 \tilde \sigma_i^2
+ {3 \over 8} \sum_{i = 1}^3 \sigma_i^2 + {3^{2 \over 3} \epsilon^2 \over 2} {1\over r^{8\over 3}} \sum_{i = 1}^3 (\sigma_i^2- \tilde \sigma_i^2) + {\cal O}\left(r^{-16/3}\right)\right] \,.
}
Using the explicit formulae \eqref{DiffForms} one can show that the leading term in the square brackets is the metric on $V_{5, 2}$ given in \cite{Bergman:2001qi}.

The $\tau = 0$ section of the $\R^4$ bundle over $S^4$ is a round four-sphere of radius
\es{RadiusIR}{
\epsilon \sqrt{K'(\epsilon^2)} = \sqrt{3 \over 2} \epsilon^{3 \over 4} \,.
}
The simplest way of showing this fact is by noting that eq.~\eqref{zToy} implies that at $\tau = 0$, we have $z_i = \bar z_i \equiv \epsilon x_i$, with $x_i$ real numbers satisfying $\sum_{i = 1}^5 x_i^2 = 1$, and eq.~\eqref{MetricDeformed} implies that
\es{BottomMetric}{
ds_8^2 = \epsilon^2 K'(\rho) \sum_{i = 1}^5 dx_i^2
= {3 \over 2} \epsilon^{3 \over 2} \sum_{i = 1}^5 dx_i^2 \,.
}
From eqs.~\eqref{ExplicitMetric}--\eqref{Gotabc} we also see that at $\tau = 0$ the metric can be written as
\es{BottomMetricAgain}{
ds_8^2 = {3 \over 2} \epsilon^{3 \over 2} \left[\nu^2 + \sum_{i = 1}^3 \sigma_i^2 \right] \,,
}
so the four-form $\nu \wedge \sigma_1 \wedge \sigma_2 \wedge \sigma_3$ is actually the volume form of a unit four-sphere \cite{Martelli:2009ga}.

\subsection{The M-theory background}

 The eleven-dimensional supergravity background \cite{Cvetic:2000db} constructed as a warped product between $\R^{2, 1}$ and the deformed cone ${\cal C}_\epsilon$ has the metric
\es{11dMetric}{
ds_{11}^2 = H^{-{2 \over 3}} dx_\mu dx^\mu + H^{1 \over 3} ds_8^2 \,,
\qquad dx_\mu dx^\mu = -(dx^0)^2 + (dx^1)^2 + (dx^2)^2 \,,
}
where $ds_8^2$ is the Calabi-Yau metric on ${\cal C}_\epsilon$, and $H$ is a function of $\tau$.  The four-form field strength $G_4$ of $11$-d supergravity and its Hodge dual $G_7 = *_{11} G_4$ are taken to be
\es{FourForm}{
G_4 &= dH^{-1} \wedge dx^0 \wedge dx^1 \wedge dx^2 + m \alpha \,, \\
G_7 &= H^2 (*_8 dH^{-1}) - m H^{-1} dx^0 \wedge dx^1 \wedge dx^2 \wedge \alpha \,,
}
where $\alpha$ is an anti self-dual closed (hence harmonic) $(2, 2)$-form on ${\cal C}_\epsilon$.   The ansatz \eqref{11dMetric}--\eqref{FourForm} is a solution to the $11$-d equations of motion for any eight-dimensional internal space and any anti self-dual $(2, 2)$-form on it provided that the function $H$ solves the equation
\es{Heom}{
\nabla^2_8 H = -{1 \over 2} m^2 \abs{\alpha}^2 \,,
}
where $\abs{\alpha}^2$ is defined through $\alpha \wedge *_8 \alpha = \abs{\alpha}^2 \vol_8$.  In the case we are interested in where the 8-d internal manifold is ${\cal C}_\epsilon$, a normalizable form $\alpha$ was found in \cite{Cvetic:2000db}:
\es{Gotalpha}{
\alpha &\equiv {3 \over \epsilon^3 \cosh^4 {\tau \over 2}} \left[a^3 c\, \nu \wedge \sigma_1 \wedge \sigma_2 \wedge \sigma_3
+ {1 \over 2}  b^3 c\, d\tau \wedge \tilde \sigma_1 \wedge \tilde \sigma_2 \wedge \tilde \sigma_3 \right] \\
{}&+ {1 \over 2 \epsilon^3 \cosh^4 {\tau\over 2}} \left[{1 \over 2} a^2 b c\, \epsilon^{ijk} d \tau \wedge \sigma_i \wedge \sigma_j \wedge \tilde \sigma_k +a b^2 c\, \epsilon^{ijk} \nu \wedge \sigma_i \wedge \tilde \sigma_j \wedge \tilde \sigma_k \right] \,.
}
In terms of the coordinates $z_i$ introduced above, its expression is given by
\es{alphaComplex}{
\alpha = {9 \sinh^4 {\tau \over 2} \over 2 \epsilon^7 \sinh^6 \tau}
\left(\epsilon_{ijklm} z_i \bar z_j dz_k \wedge dz_l \wedge d\bar z_m \right)
\wedge (z_a d\bar z_a) + \text{c.c.}
}
The solution to eq.~\eqref{Heom} can then be written as
\es{GotH}{
H =  {m^2 \over \epsilon^{{9 \over 2}}} \hat H \,,  \qquad
\hat H(\tau) = 2^{3 \over 2} 3^{11 \over 4} \int^\infty_{(2 + \cosh \tau)^{1 \over 4}} {dt \over (t^4 - 1)^{5 \over 2}} \,.
}
In obtaining \eqref{GotH} we imposed the boundary conditions that $H$ should be regular at $\tau = 0$ and that it should go to zero at large $\tau$.  Let us note that in the transverse part of the metric, $H^{1 \over 3} ds_8^2$, $\epsilon$ cancels out.  Thus, as in the KS background \cite{Klebanov:2000hb}, $\epsilon$ can be removed by rescaling the longitudinal coordinates $x^\mu$ \cite{Herzog:2002ih}.

For future reference, it is useful to note that $G_4$ can be obtained from the following three-form gauge potential $A_3$:
\es{GotA3}{
A_3 &= H^{-1} dx^0 \wedge dx^1\wedge dx^2 + m \beta \,,
}
with
\es{Gotbeta}{
\beta &= - {a c \over \epsilon^3 \cosh^4 {\tau\over 2}}
\left[ (2 a^2 + b^2) \tilde \sigma_1 \wedge \tilde \sigma_2 \wedge \tilde \sigma_3
+ {a^2 \over 2} \epsilon^{ijk} \sigma_i \wedge \sigma_j \wedge \tilde \sigma_k \right] \\
&= -{3 \over 4 \epsilon^5 \sinh^4 \tau} \epsilon_{ijklm} z_i \bar z_j dz_k \wedge dz_l \wedge dz_m
+ {9 \cosh \tau \over 8 \epsilon^5 \sinh^4 \tau} \epsilon_{ijklm} z_i \bar z_j dz_k \wedge dz_l
 \wedge d\bar z_m + \text{c.c.}
}

The number of units of M2-brane flux at fixed $\tau$ can be computed by integrating $*G_4$ over a constant $\tau$ section of ${\cal C}_\epsilon$. The result is
\es{M2BraneFlux}{
N(\tau) = {384 m^2 \Vol(V_{5, 2}) \over (2 \pi \ell_p)^6} \tanh^4 {\tau\over 2}
= {81 \pi^4 m^2 \over (2 \pi \ell_p)^6} \tanh^4 {\tau\over 2}\,,
}
where we used the fact that the volume of $V_{5, 2}$ is $27 \pi^4 /128$ \cite{Bergman:2001qi}.  Asymptotically at large $\tau$, eq.~\eqref{M2BraneFlux} becomes
\es{M2Flux}{
 N = {81 \pi^4 m^2 \over (2 \pi \ell_p)^6} \,.
}
As we discuss in more detail in section~\ref{FIELDTHEORY}, the supergravity background presented above is dual to a gauge theory where $N$ is the number of colors.

\section{Reduction to type IIA}
\label{IIAREDUCTION}

Many of the calculations in this paper will be done not using the 11-d background described above, but its dimensional reduction to type IIA string theory. Let us perform a dimensional reduction along one of the $\R^2$ directions, say $x^2$, which we take to be a circle of radius $R_{11}$.  The dimensionally-reduced background contains the following fields.  The string frame metric
\es{10dMetric}{
ds_{10}^2 = H^{-1} \left[-(dx^0)^2 + (dx^1)^2 \right] + ds_8^2
}
is a warped product of $\R^{1, 1}$ and the deformed cone ${\cal C}_\epsilon$.
The dilaton is given by
\es{NSNSOther}{
e^{2 \Phi} = H^{-1}
}
and blows up at large $\tau$, signaling that a better description of the UV physics is given by the M-theory uplift of this construction.  The NS-NS 2-form gauge potential $B_2$ and its field strength $H_3$ are
\es{NSNSB2H3}{
B_2 = H^{-1} dx^0 \wedge dx^1 \,, \qquad H_3 = dH^{-1} \wedge dx^0 \wedge dx^1 \,.
}
In the R-R sector, the two-form $F_2$ and its Hodge dual $F_8 = *_{10} F_2$ both vanish, while $F_4$ and $F_6 = *_{10} F_4$ are given by
\es{FourFormIIA}{
F_4 = m \alpha\,,  \qquad
F_6 = -m H^{-1} dx^0 \wedge dx^1 \wedge \alpha\,.
}

The string coupling $g_s$ and string length $\ell_s = \sqrt{\alpha'}$ in the type IIA theory are related to the radius of the circle we compactify over, $R_{11}$, and the Planck length $\ell_p$ in eleven dimensions through the formulae \cite{Polchinski:1998rr}
\es{11To10}{
g_s \ell_s  = R_{11} \,,
\qquad
g_s^{1\over 3} \ell_s = \ell_p \,.
}

\section{UV and IR Behavior }
\label{FIELDTHEORY}

Let us first comment on the asymptotic behavior of the geometry at large $\tau$ and its field theory interpretation.  The radial coordinate $r$ defined in \eqref{rLargetau} brings the Calabi-Yau metric on ${\cal C}_\epsilon$ to the asymptotic form of the cone over $V_{5,2}$.  The $11$-d metric (\ref{11dMetric}) asymptotes to a direct product between $AdS_4$ space with radius $L = m^{1\over 3}$ and $V_{5, 2}$ with radius $2 L$:
\es{MetricLargetau}{
ds_{11}^2 \approx {r^4 \over 16 m^{4\over 3}} dx_\mu dx^\mu + {4 m^{2 \over 3} \over r^2} dr^2 + 4 m^{2 \over 3} ds_{V_{5, 2}}^2 \ .
}
The standard AdS radial coordinates is $r_{AdS}= r^2 / (4 m^{1\over 3})$.

Except for a $\Z_2$ torsion 3-cycle, the Sasaki-Einstein space $V_{5,2}$ does not have non-trivial topology, so the 4-form field strength may asymptotically be written in terms of a well-defined
three-form gauge potential:
\es{A3Largetau}{
A_3 \approx {r_{AdS}^3 \over m^{4 \over 3}} dx^0 \wedge dx^1 \wedge dx^2
+ {m \epsilon \over r_{AdS}^{2 \over 3}} \tilde \beta \,,
}
where $\tilde \beta$ is an $SO(5)$-invariant $3$-form on $V_{5, 2}$,
 \es{tildeAlpha}{
  \tilde \beta \sim \epsilon_{ijklm} y_i \bar y_j dy_k \wedge dy_l \wedge d\bar y_m + \text{c.c.}
 }
The second term in eq.~\eqref{A3Largetau} has the interpretation of a source for a pseudoscalar operator ${\cal O}$ of conformal dimension $7/3$ or a VEV for an operator of dimension $2/3$ in the dual field theory \cite{Klebanov:1999tb}. Following \cite{Herzog:2000rz,Martelli:2009ga} we will adopt the source interpretation.\footnote{We must stress, however, that the dual dimension $7/3$ operator must be $SO(5)$ invariant because the form $\tilde \beta$ is. } Then the field theory Lagrangian is
\es{LagDeformed}{
{\cal L} = {\cal L}_{\rm CFT} + \Lambda^{2 \over 3} {\cal O} \,,
}
where $\Lambda$ is the energy scale of the relevant deformation.  From \eqref{A3Largetau} we can identify
\es{GotLambda}{
\Lambda^{2 \over 3} \sim {\epsilon \over m^{2 \over 3}}\,, \qquad \text{so} \qquad
\Lambda \sim {\epsilon^{3 \over 2} \over m}\,.
}

A crucial feature of the Stenzel space is the presence of the deformation parameter $\epsilon$.
Let us argue that the deformation is related to appearance of a VEV of a field theory operator. The leading effect of the deformation on the asymptotic metric \eqref{LargetauMetric} is the appearance of a term proportional to
 \es{AnotherOperator}{
  {1 \over r_{AdS}^{4\over 3}} \sum_{i = 1}^3 (\sigma_i^2 -\tilde \sigma_i^2)
  \sim {1 \over r_{AdS}^{4 \over 3}} \sum_{i = 1}^5 \left[(dy_i)^2 + (d\bar y_i)^2 \right]\,.
 }
Such a scaling corresponds either to a source for an operator of dimension $5/3$ or to a VEV of an operator of dimension $4/3$. We choose the latter interpretation, since usually in gauge/gravity duality a smoothing of the apex of a cone corresponds to an infrared effect where an operator gets a VEV \cite{Klebanov:1999tb}. Noting that the metric perturbation transforms as an $SO(5)$ singlet, we see that the CFT dual to $AdS_4\times V_{5,2}$ should contain an $SO(5)$-invariant scalar operator of dimension $4/3$.

For many calculations in this paper we will not need the full M-theory background and its type IIA counterpart, but only the
$\tau \rightarrow 0$ limit thereof.  Indeed, an anti-M2 brane in M-theory, or an anti-fundamental string in the type IIA reduction, placed at a non-zero value of $\tau$ will experience a force towards smaller $\tau$ and will eventually stabilize at $\tau = 0$.

At $\tau = 0$ there are a few significant simplifications.  The first is that the space ${\cal C}_\epsilon$ reduces to a round $S^4$ of radius $\sqrt{3 \over 2} \epsilon^{3 \over 4}$.  So $ds_8^2$ in \eqref{11dMetric} and \eqref{10dMetric} can be replaced by
\es{ds8IR}{
ds_8^2 \to {3 \over 2} \epsilon^{3 \over 2} d\Omega_4^2 \,,
}
where $d\Omega_4^2$ is the standard line element on a four-sphere of unit radius.  The second simplification is that $H(\tau)$ approaches a constant that can be computed from the first relation in \eqref{GotH} and
\es{GotH0}{
\hat H \to \hat H_0 = 2^{3 \over 2} 3^{11 \over 4} \int^\infty_{3^{1 \over 4}} {dt \over (t^4 - 1)^{5 \over 2}}
\approx 1.0898 \,.
}
In the warped background \eqref{11dMetric}, the radius squared of the $4$-sphere is ${3 \over 2} m^{2 \over 3} \hat H_0^{1 \over 3}$.  Lastly, the 4-form $\alpha$ becomes proportional to the volume form on $S^4$:
\es{alphaIR}{
\alpha \to {27 \over 4} \vol_{S^4} \,.
}
The number of $G_4$ flux units $\tilde M$ through the $S^4$
(or R-R four-form flux units in type IIA)
can be computed from the standard formula
\es{Mtilde}{
\tilde M = {1 \over (2 \pi \ell_p)^3} \int_{S^4} F_4 = {18 \pi^2 m \over (2 \pi \ell_p)^3} \,,
}
where $\ell_p$ is the Planck length in eleven dimensions.

Parameterizing the $S^4$ by an azimuthal angle $\psi$ and a three-sphere such that
\es{S4Param}{
d\Omega_4^2 = d \psi^2 + \sin^2 \psi\, d\Omega_3^2 \,,
}
one can write down a three-form gauge potential for the $11$-d SUGRA field $G_4 = dA_3$:
\es{A3IR}{
A_3 = {27 \over 4} m f(\psi) \vol_{S^3} \,,
}
where $\vol_{S^3}$ is the volume form on $S^3$ and the function $f(\psi)$ is given by
\es{fpsiDef}{
f(\psi) \equiv \int_0^\psi d\tilde \psi \sin^3 \tilde \psi
= {1 \over 3} \cos^3 \psi - \cos \psi + {2 \over 3} \,.
}
Similarly, one can write down the gauge potentials for $F_4 = dC_3$ and $F_6 = dC_5$ in type IIA:
\es{C3C5}{
C_3 = {27 \over 4} m f(\psi) \vol_{S^3} \,, \qquad
C_5 = -{27 \epsilon^{9\over 2}  \over 4 \hat H_0 m} f(\psi) dx^0 \wedge dx^1 \wedge \vol_{S^3} \,.
}
The gauge potentials \eqref{A3IR} and \eqref{C3C5} are well-defined everywhere except for the South pole of $S^4$, $\psi = \pi$.

It would be very interesting to understand the infrared field theory dual to this background.
Its important feature is that there are $\tilde M$ units of $G_4$ flux through the blown-up 4-sphere \cite{Martelli:2009ga}.  Similarly, after the reduction to type IIA, there are $\tilde M$ units of R-R 4-form flux through the 4-sphere. In gauge/gravity duality, the number of units of a quantized R-R flux is typically mapped to the number of colors in the gauge theory. Therefore, it is tempting to conjecture that the IR gauge theory dual to the type IIA background is a $(1+1)$-dimensional $U(\tilde M)$ SYM theory with ${\cal N}=2$ supersymmetry.  The dual gravity makes some interesting predictions about the IR properties of the gauge theory.  Adding a string at $\tau=0$ in one of the two possible orientations does not change the energy (such a string is a BPS object).  This implies that the gauge theory is not confining because separating the endpoints of the string at some large $\tau$ does not necessarily produce a linear potential. Reversing the orientation of the string creates a metastable state.

 What is the effect of adding a fundamental string in the dual $(1+1)$-dimensional $U(\tilde M)$ gauge theory? It is tempting to suggest that it is analogous to the mechanism that leads to the existence of bound states of D-strings and fundamental strings \cite{Witten:1995im}: the state with $n$ units of electric flux corresponds to adding $n$ fundamental strings. Of course, there are significant differences between the present ${\cal N}=2$ theory and the maximally supersymmetric gauge theory studied in \cite{Witten:1995im}. In particular, we would need to show why the addition of electric flux in one direction preserves supersymmetry, and in the other creates a metastable state.

\subsection{BPS domain walls}
\label{BPS}

In the M-theory background, an M5-brane wrapped over the $S^4$ is a BPS domain wall. Since there are $\tilde M$ units of $G_4$ flux through the $S^4$, this domain wall interpolates between the branches of the moduli space containing no space-time filling M2-branes and $\tilde M$ space-time filling M2-branes.\footnote{Such a domain wall is analogous to the NS5-brane wrapped over the $S^3$ at the bottom of the warped deformed conifold; this BPS domain wall interpolates between vacua with no D3-branes and $M$ D3-branes \cite{Kachru:2002gs,Dymarsky:2005xt}.}
Similarly, in the dimensionally reduced type IIA theory, there exists a BPS domain wall which is a D4-brane wrapped over the $S^4$; it interpolates between the branch with no BPS fundamental strings and with $\tilde M$ BPS fundamental strings.

Let us construct this domain wall as a solution  
in the D4-brane world volume field theory. The D4-brane action is
\es{D4BraneAction}{
S &= -\mu_4 \int d^5 x\, e^{-\Phi} \sqrt{-\det (g_{ab} + B_{ab} + 2 \pi \alpha' {\cal F}_{ab})}
+ \mu_4 \int C_5 \\
{}&+ \mu_4 \int C_3 \wedge (B_2 + 2 \pi \alpha' {\cal F}) \,,
}
where $\mu_4$ is the D4-brane tension
\es{mu4Def}{
\mu_4 = {2 \pi \over g_s (2 \pi \ell_s)^5} = {1 \over (2 \pi)^4 g_s (\alpha')^{5 \over 2}} \,,
}
and ${\cal F} = d{\cal A}$ is the two-form field strength on the brane.  Note that since $F_6 = -B_2 \wedge F_4$ for our background (see eq.~\eqref{FourFormIIA}), one can choose the gauge potentials $C_3$ and $C_5$ so that $C_5 + C_3 \wedge B_2 = 0$.  The Chern-Simons term in the D4-brane action therefore reduces to the integral of $C_3 \wedge {\cal F}$ over the brane world-volume.

We are interested in the case where our D4-brane spans $(x^0, x^1)$ and wraps the $S^3$ inside the $S^4$ located at $\tau=0$.  Let's take ${\cal F} = 0$.  The Lagrangian for the azimuthal angle $\psi$ of the $S^4$ is found to be
\es{Lagjust}{
{\cal L} &=  -\tilde M V_{\rm string}^{(0)} \left[ \sqrt{1 + {3 \hat H_0 m^2 \over 2 \epsilon^3}  (\partial_\mu \psi)^2} \sqrt{{\hat H_0 \over 96} \sin^6 \psi +  {9 \over 64} f(\psi)^2   }  -{3 \over 8} f(\psi) \right] \,.
}
The domain wall  
is a solution $\psi(x^1)$ which interpolates between $\psi=\pi$ at $x^1 = -\infty$ and $\psi=0$ at some $x^1 = x^1_0$. The infinite extension of the D4-brane towards $x^1=-\infty$ corresponds to
$\tilde M$ fundamental strings emanating from it \cite{Callan:1998iq,Callan:1999zf}.

For fields depending on $x^1$ only, it is convenient to use a Hamiltonian formalism  and define the canonical momentum $P_\psi = {\partial \cal L} / \partial (\partial_1 \psi)$.  The corresponding Hamiltonian density ${\cal H} \equiv P_\psi \partial_1 \psi - {\cal L}$ is
\es{HamiltDensity}{
{\cal H} &=  \tilde M V_{\rm string}^{(0)} \left[ \sqrt{{\hat H_0 \over 96} \sin^6 \psi +  {9 \over 64} f(\psi)^2  - {32 \pi^4 \hat H_0 g_s^2 \alpha'{}^5 \over 243 \epsilon^6} P_\psi^2  }  -{3 \over 8} f(\psi)  \right] \,.
}
Since the Lagrangian density \eqref{Lag} does not depend on $x^1$, the Hamiltonian \eqref{HamiltDensity} is constant on all solutions to the Hamilton equations following from it.
Since we want the solution to asymptote to $\psi = \pi$ as $x^1 \to -\infty$, we necessarily have $P_\psi(-\infty) = 0$ and, as a consequence, $ {\cal H} = 0$ also (see~\cite{Kachru:2002gs} for a similar argument).  This last requirement allows us to find the trajectory of the solution in phase space:
\es{PpsiRelation}{
P_\psi = \pm { 9 \epsilon^3 \over 32 \pi^2 g_s \alpha'{}^{5 \over 2}} \sin^3 \psi \,.
}
The Hamilton equation for $\psi'$ (where the prime denotes derivative with respect to $x^1$) implies
\es{HamiltonEqBPS}{
\psi' = \pm {2 \epsilon^{3 \over 2} \over 9 m}  {\sin^3 \psi \over f(\psi)} \,,
}
which can be integrated to give
\es{BPSDomainWall}{
x^1 - x^1_0 = \mp {3 m \over 4 \epsilon^{3 \over 2}} \left[-\tan^2 {\psi \over 2} + \log \cos^4 {\psi \over 2} \right] \,,
}
for an integration constant $x^1_0$.  Note that $\psi$ approaches $\pi$ as $x^1 \to \pm \infty$, while $\psi = 0$ at $x^1 = x^1_0$.  Since the Hamiltonian vanishes for this solution, the on-shell action is given just by the phase space area:
\es{SOnShellBPS}{
S_{\rm D4} = \int dx^0 \int_0^\pi d\psi \, P_\psi \,.
}
Identifying $S_{\rm D4} = \int dx^0\, m_{\rm D4}$ where $m_{\rm D4}$ is the mass of the BPS domain wall, we obtain
\es{WallTension}{
m_{\rm D4} = {3 \epsilon^3 \over 8 \pi^2 g_s \alpha'{}^{5 \over 2}}  \,.
}
This formula agrees precisely with the mass of a static D4-brane wrapping the $S^4$ at $\tau = 0$ and located at a fixed value of $x^1$.

The domain wall \eqref{BPSDomainWall} can be interpreted as follows.  Because of the $\tilde M$ units of $F_4$-flux through the $S^4$ the D4-brane wraps, the D4-brane sources the open string gauge field ${\cal A}$ creating $\tilde M$ fundamental strings ending on it.  The strings ending on the D4-brane pull on the brane deforming it and making it extend over a range of $x^1$ \cite{Callan:1998iq,Callan:1999zf}.  The actual BPS D4-brane gets deformed into a domain wall that starts at $x^1_0$, where $\psi = 0$, and then continues to, say, smaller values of $x^1$, as $\psi(x^1)$ varies from $0$ to $\pi$; $\psi(x^1)$ reaches $\pi$ only asymptotically as $x^1 \to -\infty$.  Thus, the $\tilde M$ fundamental strings that end on the brane thicken into a (hollow) tube---and indeed, one can check that the asymptotic tension of the tube (as computed from the DBI term in the action) equals precisely the tension of $\tilde M$ fundamental strings.  The $\tilde M$ fundamental strings, however, don't cost any energy because they are BPS objects. This explains why the mass of the BPS domain wall agrees with that of a D4-brane wrapping the $S^4$ at the bottom of the throat at a fixed value of $x^1$.

\section{Metastable states from string and M-theory perspective}

When an anti-M2 brane filling the $(x^0, x^1, x^2)$ directions, and located at fixed values of the other coordinates, is placed in the M-theory background \eqref{11dMetric}--\eqref{FourForm}, it falls towards smaller values of $\tau$ until it stabilizes at $\tau = 0$.   So let us examine a stack of $p$ anti-M2 branes at $\tau = 0$ located at a fixed point on $S^4$, which without loss of generality can be assumed to be the North pole, $\psi = 0$.  Because of the non-zero $A_3$ field, one expects the anti-M2 branes to polarize through an analog of the Myers effect \cite{Myers:1999ps,Bachas:2000ik}, and get blown up into an M5-brane that fills the $(x^0, x^1, x^2)$ directions that the original anti-M2 brane was filling, and in addition wraps an $S^3$.  The most likely scenario is that the M5-brane is located at $\tau = 0$ and wraps an azimuthal $S^3 \subset S^4$ at a fixed value of $\psi$.\footnote{Another possibility is, for instance, that the M5-brane wraps an $S^3 \subset \R^4$ at a fixed value of $\tau$.  In Appendix~\ref{BLOWUP} we show that the potential for such an M5-brane has only one minimum at $\tau = 0$, so there is no tendency for the anti-M2 branes to blow up into an M5-brane located at fixed $\tau>0$.}

To demonstrate this effect, one would need to compute the potential for an M5-brane with $p$ anti-M2 branes dissolved in it, and see whether it has a local minimum at some value of $\psi > 0$.  We find it more convenient, however, to study this process from the point of view of the dimensionally reduced type IIA theory presented in section~\ref{IIAREDUCTION}. Since the dimensional reduction to type IIA was performed along a direction parallel to the branes, the anti-M2 branes reduce to anti-fundamental strings in type IIA, while the M5-brane becomes a D4-brane. If we then place a number of anti-fundamental strings at $\tau = \psi = 0$, we want to know whether they blow up into a D4-brane filling $(x^0, x^1)$ and an azimuthal $S^3 \subset S^4$ at some fixed $\psi$ and $\tau = 0$.  We thus compute the potential for such a D4-brane as a function of $\psi$.  Clearly, the corresponding potential for an M5-brane in M-theory can be inferred from that of a D4-brane in type IIA.

Let's first get some idea of the energy scales involved.  From the action for $Q_{\rm NS1}$ fundamental strings,
\es{FundStringAction}{
S = -{\abs{Q_{\rm NS1}} \over 2 \pi \alpha'} \int d^2 x \sqrt{-g} + {Q_{\rm NS1} \over 2 \pi \alpha'} \int B_2 \,,
}
we notice that if we put any number $Q_{\rm NS1}>0$ of fundamental strings at $\tau = \psi = 0$, there is an exact cancellation between the two terms in \eqref{FundStringAction}, indicating that fundamental strings at $\tau = 0$ are BPS objects.  If, however, one places some number $p = -Q_{\rm NS1} >0$ of anti-fundamental string at $\tau = \psi = 0$, the two terms in \eqref{FundStringAction} are equal and they add.  The potential $V_{\rm string}^{(0)}$ per unit $x^1$ length for just one such anti-fundamental string ($p = 1$) is then
\es{Vstring}{
V_{\rm string}^{(0)} = {1 \over \pi \alpha'} {\epsilon^{9 \over 2} \over \hat H_0 m^2} \,.
}
In the M-theory uplift of this construction, M2-branes placed at $\tau = \psi = 0$ are also BPS, while anti-M2 branes are not.  The potential $V_{\rm M2}^{(0)}$ per unit $(x^1, x^2)$-area for one anti-M2 should be identified with $V_{\rm string}^{(0)} / (2 \pi R_{11})$, so
\es{VM2}{
V_{\rm M2}^{(0)} =  {1 \over 2 \pi^2 \ell_p^3} {\epsilon^{9 \over 2} \over \hat H_0 m^2}
}
upon using \eqref{11To10}.

\subsection{Fundamental strings blowing up into a D4-brane}
\label{POTENTIAL}

We can add $p$ units of anti-fundamental string charge dissolved in the D4-brane by turning on the electric component ${\cal F}_{01}$ of the world-volume flux on the brane \cite{Herzog:2002ss}.  It is convenient to define a rescaled electric field ${\cal E}$ by
\es{F01Normalization}{
{\cal F}_{01} = {1 \over 2 \pi \alpha'} {\epsilon^{9\over 2} \over \hat H_0 m^2} {\cal E} \,.
}
Using eqs.~\eqref{10dMetric}--\eqref{FourFormIIA}, \eqref{ds8IR}--\eqref{alphaIR}, and \eqref{C3C5} that describe the IR limit of the type IIA background, we can write the D4-brane action \eqref{D4BraneAction} as
\es{EffectiveLagSimp}{
S = \int d^2 x \, {\cal L}_{\cal E} \,, \qquad
{\cal L}_{\cal E} \equiv - A_1 \sin^3 \psi \sqrt{1 - (1 + {\cal E})^2} - A_2 f(\psi) {\cal E} \,,
}
where we have defined
\es{ABDefs}{
A_1 \equiv \mu_4 {3^{3 \over 2} \epsilon^{9 \over 2} \over 2^{3 \over 2} \hat H_0^{1\over 2} m} \Vol(S^3) \,,
\qquad
A_2 \equiv \mu_4 {27 \epsilon^{9 \over 2} \over 4 \hat H_0 m}  \Vol(S^3) \,,
}
and chosen the D4-brane orientation that gives the minus sign in the last term of \eqref{EffectiveLagSimp}.

The electric field ${\cal F}_{01}$ defined in \eqref{F01Normalization} is in general not a conserved quantity in the sense that it may depend on $x^0$ and $x^1$.  Indeed, in the gauge where ${\cal A}_0 = 0$, the electric field ${\cal F}_{01} = \partial_0 {\cal A}_1$ has the form of a velocity field, and it is usually the momentum, not the velocity, that is a conserved quantity.  As in any electrostatics problem, the conserved quantity in the absence of sources is the electric displacement $\partial {\cal L} / \partial {\cal F}_{01}$, which in our case equals the fundamental string charge $-p$ \cite{Callan:1995xx}.
We can express the Lagrangian ${\cal L}_{\cal E}$ in terms of $p$ by performing a Legendre transform:
\es{LagD}{
{\cal L}_{\cal D} = {\cal L}_{\cal E} - {\cal D} {\cal E} = -\sqrt{A_1^2 \sin^6 \psi + \left[{\cal D} + A_2 f(\psi)\right]^2 }
+ \left[{\cal D} + A_2 f(\psi) \right]\,,
}
where the rescaled displacement ${\cal D} = \partial {\cal L}_{\cal E}  / \partial {\cal E}$ is related to $p$ through
\es{PQuantization}{
 {\cal D} = {1 \over 2 \pi \alpha'} {\epsilon^{9\over 2} \over \hat H_0 m^2} (-p) \,.
}
This Lagrangian should be identified with minus the potential of the D4-brane per unit $x^1$-length, ${\cal L}_{\cal D} = -V(\psi)$.  Using \eqref{Mtilde} as well as \eqref{ABDefs}--\eqref{PQuantization}, one finds
\es{LagDSimp}{
V(\psi) &=  \tilde M V_{\rm string}^{(0)} \left[\sqrt{{\hat H_0 \over 96} \sin^6 \psi +  \left(  {3 \over 8} f(\psi) - {p \over 2 \tilde M}\right)^2   }  -{3 \over 8} f(\psi) + {p \over 2 \tilde M} \right] \,.
}

Let us try to understand a few limits of this formula.  First, when $\psi = 0$ we have $f(0) = 0$, and therefore $V(0)$ vanishes when $p<0$, and $V(0) = \abs{p} V_{\rm string}^{(0)}$ when $p>0$.  This is just a consistency check that our D4-brane has $(-p)$ units of fundamental string charge, and when the $S^3$ it wraps shrinks to zero size its energy is precisely that of the fundamental (or anti-fundamental) strings dissolved in it---see the discussion around eq.~\eqref{Vstring}. Secondly, when $\psi = \pi$, we have $f(\pi) = {4 \over 3}$, so as long as $p<\tilde M$ the potential $V(\pi)$ vanishes, suggesting that at $\psi = \pi$ the D4-brane represents a BPS object.  Indeed, as we move the D4-brane from $\psi = 0$ to $\psi = \pi$ we are effectively inducing $\tilde M$ extra units of fundamental string charge, so at $\psi = \pi$ we are describing $\tilde M - p$ fundamental strings.  If $p<\tilde M$ these fundamental strings are BPS and cost no energy; if $p>\tilde M$ they are in fact anti-strings and $V(\pi)>0$.

\begin {figure}
\center\includegraphics [width=0.7\textwidth]{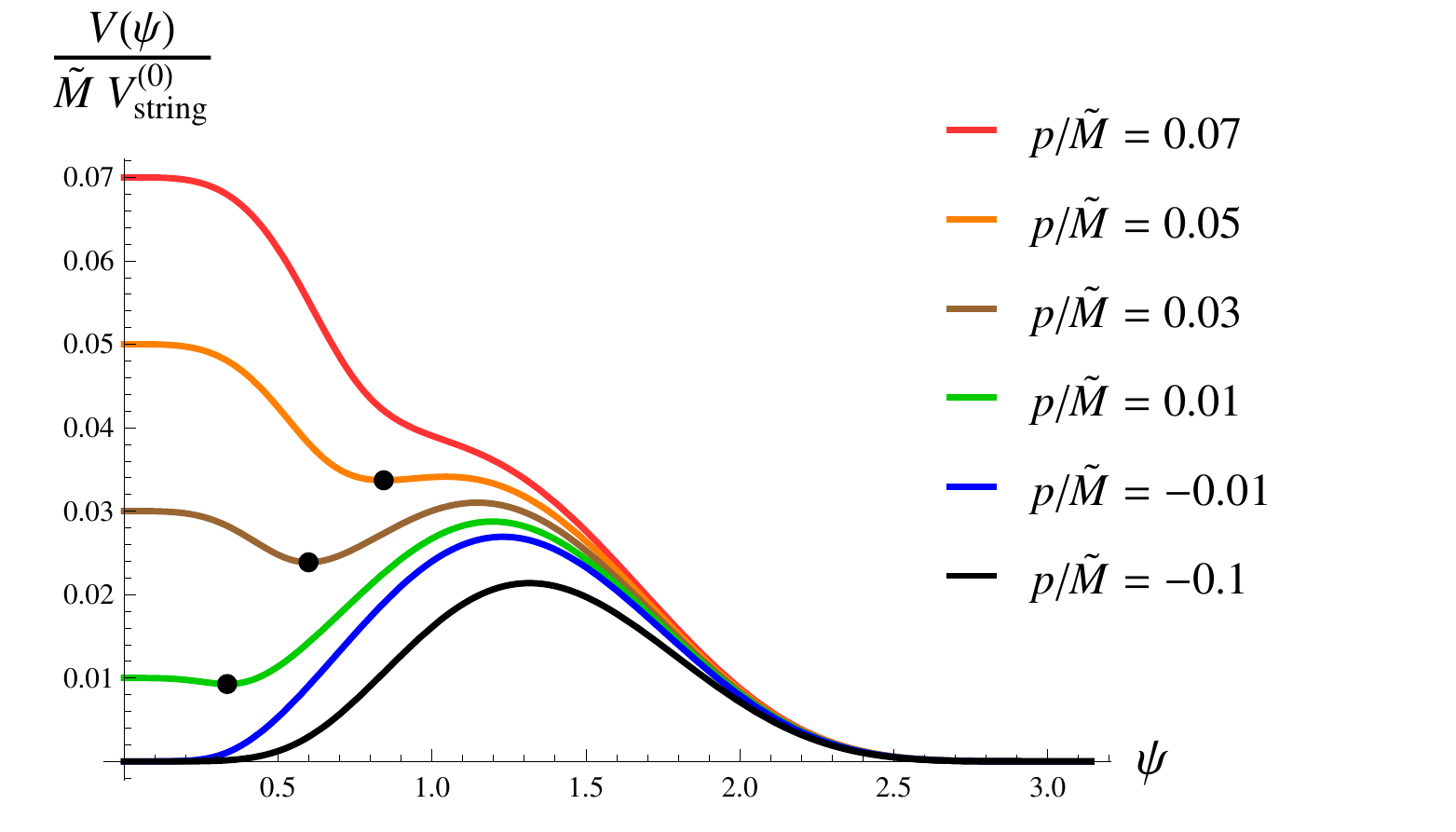}
\caption {The D4-brane potential given in eq.~\eqref{LagDSimp}.  The potential has a metastable minimum marked by a black dot for $p/\tilde M \lesssim 0.0538$---see eq.~\eqref{GotpMax}. \label {fig:D4Potential}}
\end {figure}

In figure~\ref{fig:D4Potential} we plotted $V(\psi)$ for various values of $p/\tilde M$.
This plot shows that for small positive values of $p/\tilde M$ (corresponding to a small number of anti-strings) the potential has a metastable minimum at some $\psi = \psi_{\rm min}$, the global minimum being at $\psi = \pi$ as discussed above.  The metastable minimum disappears above $p/\tilde M \approx 0.0538$ where $V(\psi)$ becomes a monotonically decreasing function of $\psi$, or for $p/\tilde M \leq 0$ where $V(\psi)$ has two supersymmetric minima at $\psi = 0$ and $\psi = \pi$ and is strictly positive for all other values of $\psi$.

\begin {figure}
\center\includegraphics [width=0.5\textwidth]{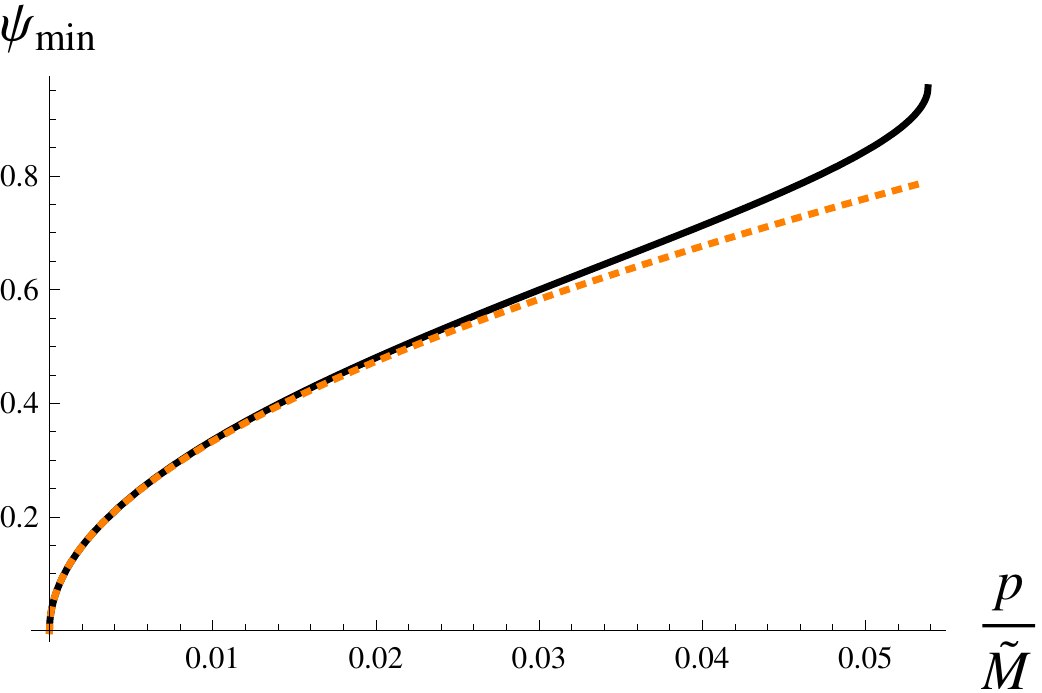}
\caption {The angle $\psi_{\rm min}$ at which there is a metastable minimum as a function of $p/\tilde M$.  The exact solution value of $\psi_{\rm min}$ found by solving eq.~\eqref{QuarticEquation} is plotted in solid black, and the small $p$ approximation \eqref{SmallpBehavior} is plotted in dotted orange.  \label {fig:psiminPlot}}
\end {figure}

It is possible to get some analytic insight into the location of the metastable minimum and its vacuum energy.  Using \eqref{LagDSimp} one can show that $\cos \psi_{\rm min}$ satisfies the following quartic equation
\es{QuarticEquation}{
(3 - 2 \hat H_0) \cos^4 \psi_{\rm min} - 2 (6 - \hat H_0) \cos^2 \psi_{\rm min}
+ 12 \left(1 - 2 {p \over \tilde M} \right) \cos \psi_{\rm min} - 3 = 0 \,.
}
This equation does have closed form solutions, but their expressions are long and not very illuminating.  Only one of these solutions corresponds to the metastable minimum and we plotted it against $p/\tilde M$ in figure~\ref{fig:psiminPlot}.
At small $p/\tilde M$ one finds that
\es{SmallpBehavior}{
\cos \psi_{\rm min} = 1 - {6 \over \hat H_0} {p \over \tilde M} + {\cal O}(p^2/\tilde M^2) \,,
}
and then from \eqref{LagDSimp},
\es{VpsiSmallp}{
V(\psi_{\rm min}) = \tilde M V_{\rm string}^{(0)} \left[{p \over \tilde M} -  {9 \over \hat H_0^2}
{p^2 \over \tilde M^2} + {\cal O}(p^3/\tilde M^3) \right]\,.
}
As can be seen from figure~\ref{fig:psiminPlot}, eq.~\eqref{SmallpBehavior} approximates quite well the exact solution up to $p/\tilde M \approx 0.03$.

It is possible to use eq.~\eqref{QuarticEquation} to find an analytic formula for the maximal value of $p/\tilde M$ at which the metastable minimum disappears.  At this maximal value, the potential has an inflection point with zero slope at $\psi_{\rm min}$, so the derivative of the function appearing on the LHS of eq.~\eqref{QuarticEquation} also vanishes:
\es{DerivativeVanishes}{
4 (3 - 2 \hat H_0) \cos^3 \psi_{\rm min} - 4 (6 - \hat H_0) \cos \psi_{\rm min} + 12 \left( 1 - 2 {p_{\rm max} \over \tilde M}\right) = 0 \,.
}
Combining eqs.~\eqref{QuarticEquation} and \eqref{DerivativeVanishes}, we obtain $\psi_{\rm min} = \arccos {1 \over \sqrt{3}} \approx 0.955$.  This gives
\es{GotpMax}{
{p_{\rm max} \over \tilde M} = {1 \over 2} - {15 - \hat H_0 \over 18 \sqrt{3}} \approx 0.0538.
}

As a last comment, we note that we could have also done the above computation in the M-theory uplift, where the D4-brane wrapped over an azimuthal $S^3$ becomes an M5-brane wrapped over the same $S^3$ filling now the directions $(x^0, x^1, x^2)$ instead of just $(x^0, x^1)$.  The potential density for this M5-brane is
\es{Vtilde}{
\tilde V(\psi) = {V(\psi) \over 2 \pi R_{11} } \,,
}
so the curves in figure~\ref{fig:D4Potential} also represent $\tilde V(\psi) / (\tilde M V_{\rm M2}^{(0)})$, as can be noted from the fact that the energy density $V_{\rm M2}^{(0)}$ of an M2-brane at $\psi = \tau = 0$ defined in \eqref{VM2} equals $V_{\rm string}^{(0)} / (2 \pi R_{11})$.

\section{The decay of the false vacuum}

Over long enough time scales, the metastable vacuum described in the previous section undergoes quantum tunneling to the  true vacuum; this is the supersymmetric state where, in the type IIA case, the D4-brane is at the South pole of $S^4$ and reduces to $\tilde M-p$ fundamental strings.  The tunneling event consists of the nucleation of a ``bubble'' of true vacuum, namely a configuration where $\psi$ approaches $\psi_{\rm min}$ as $x^1 \to \pm \infty$, while close to the center of the bubble $\psi$ is on the other side of the potential barrier from figure~\ref{fig:D4Potential}.  This configuration then evolves in time classically:  the outward pressure makes the bubble expand and, eventually, the field $\psi$ will be in the true vacuum at all values of $x^1$.

In order to compute the tunneling rate, we need to consider a generalization of the Lagrangian discussed in the previous section that allows for the possibility of making $\psi$ depend on $x^0$ and $x^1$.  Such a generalization is
\es{Lag}{
{\cal L} &=  -\tilde M V_{\rm string}^{(0)} \left[ \sqrt{1 + {3 \hat H_0 m^2 \over 2 \epsilon^3}  (\partial_\mu \psi)^2} \sqrt{{\hat H_0 \over 96} \sin^6 \psi +  \left( {3 \over 8} f(\psi) - {p \over 2 \tilde M}\right)^2   }  -{3 \over 8} f(\psi) +  {p \over 2 \tilde M} \right] \,,
}
where $(\partial_\mu \psi)^2 \equiv -(\partial_0 \psi)^2 + (\partial_1 \psi)^2$.  Of course, when $\psi$ is a constant, this Lagrangian reduces to minus the potential $V(\psi)$ studied in the previous section; see eq.~\eqref{LagDSimp}.

In the rest of this section we calculate the tunneling rate from the metastable state to the true vacuum.   As explained in \cite{Coleman:1977py}, the decay of the false vacuum is mediated by an $O(2)$-invariant Euclidean bounce solution where $\psi$ depends on the Euclidean 2-d radius $r \equiv \sqrt{(x^0)^2 + (x^1)^2}$ and $\psi \to \psi_{\rm min}$ as $r \to \infty$. At $r = 0$ $\psi$ is on the other side of the potential barrier.  The action of the bounce can be used to compute the tunneling rate per unit volume, $\Gamma/V= A e^{-B}$. In particular, the tunneling coefficient $B$ equals the on-shell action of the bounce.  In section~\ref{SMALLP} we give an estimate of this tunneling coefficient based on a small $p$ approximation.   In section~\ref{TUNNELING} we compute $B$ numerically at all values of $p < p_{\rm max}$.  Finally, in section~\ref{MTHEORY} we redo these calculations in M-theory uplift of the type IIA background.

\subsection{Small $p$ approximation}
\label{SMALLP}

As explained above, the decay of the false vacuum is realized through the nucleation of a bubble of true vacuum, which is nothing but a D4-$\overline{\rm D4}$ pair with $p$ anti-strings on either side of the D4-branes and $\tilde M-p$ strings in between the branes.  For $p$ small, the metastable state of $p$ strings is close to $\psi = 0$, and their energy in the metastable state is approximately $p V_{\rm string}^{(0)}$, as can be seen from eq.~\eqref{VpsiSmallp}.  Since the $\tilde M-p$ strings in between the two branes do not cost any energy, the energy gained by nucleating the two branes is
\es{EnergyD4D4bar}{
E = -p V_{\rm string}^{(0)} (x_2 - x_1) + \sqrt{m_{\rm D4}^2 + P_1^2} + \sqrt{m_{\rm D4}^2 + P_2^2} \,,
}
where $m_{\rm D4}$ is the effective $(1+1)$-d mass of the wrapped branes, and $x_i$ and $P_i$, $i = 1, 2$, are their positions and momenta.  At small $p$, the quantity $m_{\rm D4}$ is well-approximated by 
(\ref{WallTension}).  In this limit, the decay of the false vacuum is nothing but Schwinger pair-production in the presence of a constant external electric field.

Standard quantum mechanics formulae for the trajectory $x_2 = -x_1$ with $E = 0$ give the tunneling coefficient in the WKB approximation:
\es{TunnelingRate}{
B = 4 \int_{0}^{x_1^*} dx_1\, \abs{P_1(x_1)} = {\pi m_{\rm D4}^2 \over p V_{\rm string}^{(0)}}\,,
}
where $x_1^* = {m_{\rm D4} / (p V_{\rm string}^{(0)}})$.  At small $p$ we therefore expect $B \sim 1/p$, so the lifetime of the metastable state can be made arbitrarily large by taking $p$ to be sufficiently small compared to $\tilde M$.

The same result can be obtain in Coleman's formalism \cite{Coleman:1977py} that is based on finding a Euclidean bounce with $O(2)$ symmetry that satisfies the boundary condition that $\psi \to \psi_{\rm min}$ at large Euclidean radius.  When $p$ is small, this bounce looks like a domain wall at a large radius $r_*$; for $r\gg r_*$, $\psi \approx \psi_{\rm min} \approx 0$, while for $r \ll r_*$, we have $\psi \approx \pi$.  Since $r_*$ is large in this limit, the tension of this domain wall can be approximated by $m_{\rm D4}$.  The difference in the Euclidean action of this configuration and that of the configuration where $\psi = \psi_{\rm min}$ everywhere is
\es{EuclideanActionRStar}{
S_E(r_*) = -\pi r_*^2 p V_{\rm string}^{(0)} + 2 \pi r_* m_{\rm D4} \,.
}
The solution to the Euclidean equations of motion will have minimal action, so $dS_E/ dr_* = 0$, giving
\es{GotrStar}{
r_* = {m_{\rm D4} \over p V_{\rm string}^{(0)}} \qquad
\Longrightarrow
\qquad
S_E = {\pi m_{\rm D4}^2 \over p V_{\rm string}^{(0)}} \,.
}
According to \cite{Coleman:1977py} the tunneling coefficient $B$ is precisely equal to the Euclidean action of the bounce that mediates the transition, and it can be seen that the value of $B = S_E$ computed in \eqref{GotrStar} agrees with the one in \eqref{TunnelingRate}.

\subsection{Tunneling rate from a smooth Euclidean bounce}
\label{TUNNELING}

When $p$ is not necessarily small, in order to compute the tunneling rate, we proceed along the lines of \cite{Coleman:1977py} and find the Euclidean bounce that mediates the false vacuum decay.  This Euclidean solution is an extremum---but not a minimum---of the Euclidean action
\es{EuclideanAction}{
S_E = -\int d^2 x\, {\cal L}_E \,,
}
where ${\cal L}_E$ is the Euclidean Lagrangian computed from \eqref{Lag} by substituting $(\partial_\mu \psi)^2 = (\partial_0 \psi)^2 + (\partial_1 \psi)^2$.  For an $O(2)$-invariant bounce where $\psi$ depends only on the Euclidean radius $r \equiv \sqrt{(x^0)^2 + (x^1)^2}$ the action becomes
\es{SERadial}{
S_E =  \tilde M V_{\rm string}^{(0)} \int dr\, 2 \pi r \left[ \sqrt{1 + {3 \hat H_0 m^2 \over 2 \epsilon^3}\psi'^2 } \sqrt{{\hat H_0 \over 96} \sin^6 \psi +  \left(  {3 \over 8} f(\psi) - {p \over 2 \tilde M}\right)^2   }  -{3 \over 8} f(\psi) +  {p \over 2 \tilde M} \right]
}
where $\psi' \equiv {d \psi / d r}$.  It is convenient to rescale the radial variable $r$ so that the first term in the square brackets contains a factor of $\sqrt{1 + \psi'^2}$.  After such a rescaling, a little algebra gives
\es{SERadialAgain}{
S_E = {\pi m_{\rm D4}^2 \over \tilde M V_{\rm string}^{(0)}}
  \int dr\, {108\, r\over \hat H_0} \left[ \sqrt{1 + \psi'^2 } \sqrt{{\hat H_0 \over 96} \sin^6 \psi +  \left(  {3 \over 8} f(\psi) - {p \over 2 \tilde M}\right)^2   }  -{3 \over 8} f(\psi) +  {p \over 2 \tilde M} \right] \,.
}

\begin {figure}
\center\includegraphics [width=0.6\textwidth]{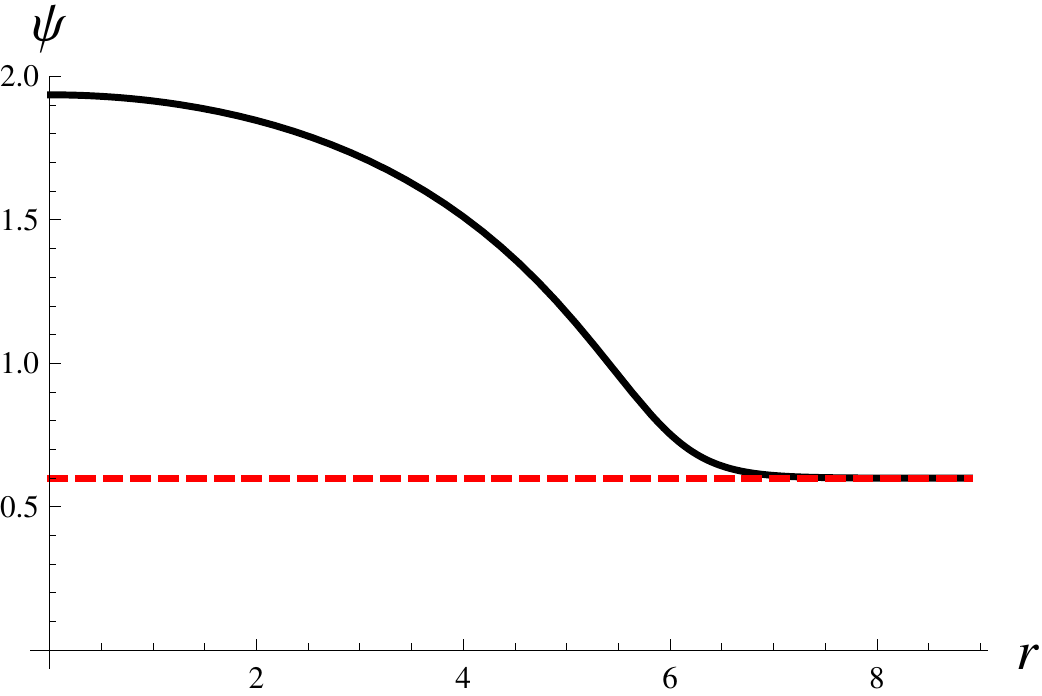}
\caption {A sample solution to the equation of motion following from \eqref{SERadialAgain} for the case $p/\tilde M = 3/100$.  The numerical solution is shown in black, and the asymptotic value it reaches as $r \to \infty$ is shown in dashed red. \label {fig:SampleDomainWall}}
\end {figure}

\begin {figure}
\center\includegraphics [width=0.6\textwidth]{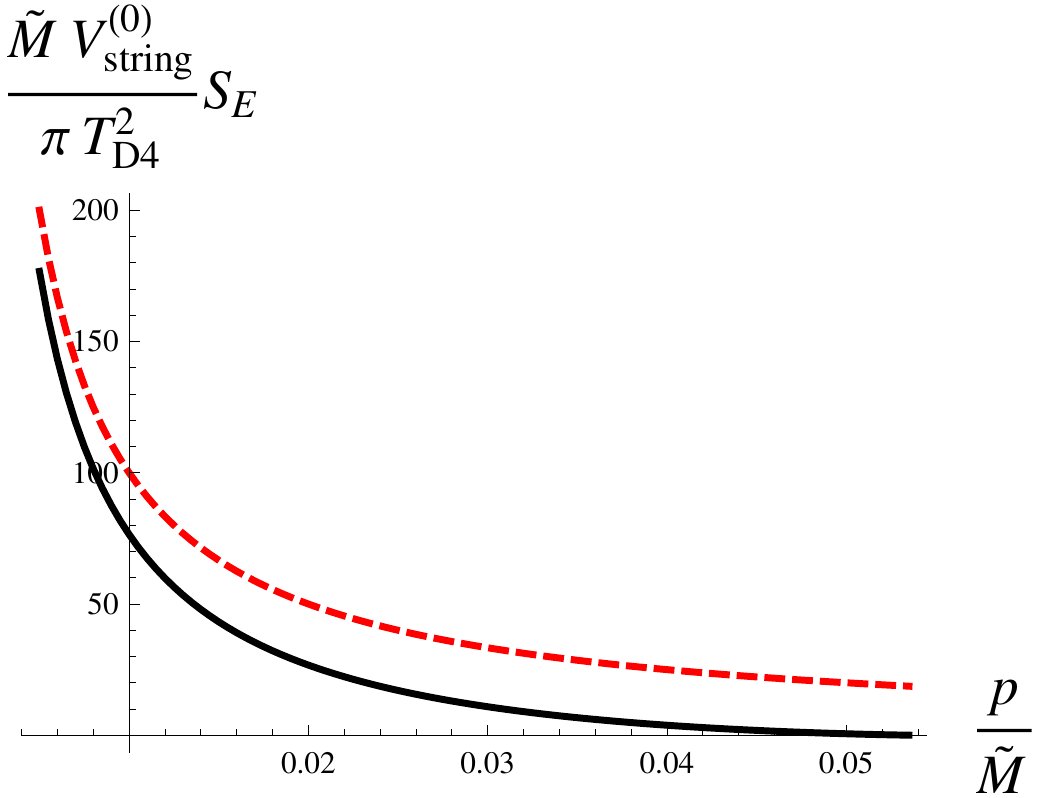}
\caption {The on-shell action for the Euclidean bounce that mediates the decay of the metastable vacuum in the type IIA construction.  The numerical results are shown in black, and the small $p$ approximation \eqref{GotrStar} is shown in dashed red.
\label {fig:CompareIIA}}
\end {figure}

We solved the Euler-Lagrange equations following from \eqref{SERadialAgain} numerically with the boundary conditions that 
$\psi'(0)=0$ and that $\psi(r)$ should approach $\psi_{\rm min}$ at large $r$.  We plot such a solution in figure~\ref{fig:SampleDomainWall}.
In figure~\ref{fig:CompareIIA} we show the value of $S_E$ as a function of $p/\tilde M$, and we compare it to the small $p$ approximation \eqref{GotrStar}.

\subsection{False vacuum decay in M-theory}
\label{MTHEORY}

We can also consider the decay of the false vacuum in the M-theory uplift of our type IIA background.  Of course, when the eleventh direction (which we have taken to be $x^2$) is compactified on a {\em small} circle of radius $R_{11}$, the tunneling rate is the same as in type IIA.  But what happens if we don't compactify the $x^2$ direction?
 Strictly speaking, we need to start by considering the action of an M5-brane with M2-brane charge dissolved in it.  However, one can argue that because in the limit where $x^2$ is compactified on a small circle this action should reduce to that of a D4-brane, the effective $(2+1)$-d Lagrangian density of an M5-brane with $(-p)$ units of M2-brane charge is just $1/(2 \pi R_{11})$ times the $(1 + 1)$-d Lagrangian in \eqref{Lag}:
\es{LagM}{
\tilde {\cal L} &=  -\tilde M V_{\rm M2}^{(0)} \left[ \sqrt{1 + {3 \hat H_0 m^2 \over 2 \epsilon^3}  (\partial_\mu \psi)^2} \sqrt{{\hat H_0 \over 96} \sin^6 \psi + \left(  {3 \over 8} f(\psi) - {p \over 2 \tilde M}\right)^2   }  -{3 \over 8} f(\psi) +  {p \over 2 \tilde M} \right] \,.
}
Here, by Lorentz invariance we have $(\partial_\mu \psi)^2 \equiv -(\partial_0 \psi)^2 + (\partial_1 \psi)^2 + (\partial_2 \psi)^2$, and we expressed the answer in terms of the energy density of an anti-M2 brane $V_{\rm M2}^{(0)}$ at $\psi = \tau = 0$ that was defined in \eqref{VM2}.

The Euclidean bounce that mediates the decay has $O(3)$ symmetry in this case.  In the small $p$ limit, this solution looks like a spherically-symmetric domain wall at some fixed value of the Euclidean radius $r_*$, where $\psi$ is approximately equal to $\psi_{\rm min} \approx 0$ for small Euclidean radius $r \ll r_*$, and approximately equal to $\pi$ for $r \gg r_*$.  The tension of this domain wall is in this limit well approximated by the tension $T_{\rm M5}$ of the BPS M5-brane wall obtained as an uplift of the construction in section~\ref{BPS}:
\es{GotTM5}{
T_{\rm M5} = {m_{\rm D4} \over 2 \pi R_{11} } = {3 \epsilon^3 \over 16 \pi^3 \ell_p^6} \,.
}
The on-shell action for a domain wall at some $r_*$ is then
\es{SESmallpM}{
S_E(r_*) = - {4 \pi r_*^3 \over 3} p V_{\rm M2}^{(0)} + 4 \pi r_*^2 T_{\rm M5}  \,,
}
where $V_{\rm M2}^{(0)}$ was defined in \eqref{VM2}.  Extremizing \eqref{SESmallpM} with respect to $r_*$, we get
\es{GotrStarM}{
r_* = {2 T_{\rm M5} \over p V_{\rm M2}^{(0)} } \qquad
\Longrightarrow
\qquad
S_E =  {16 \pi  T_{\rm M5}^3 \over 3 p^2 \left( V_{\rm M2}^{(0)}\right)^2 }
 = {4 \hat H_0^2 \over 729} { \tilde M^4 \over p^2} \,.
}
The tunneling coefficient $B = S_E$ behaves as $1/p^2$ at small $p$ in this case, so again the lifetime of the metastable state can be made arbitrarily large by taking $p/\tilde M$ to be sufficiently small.

\begin {figure}
\center\includegraphics [width=0.6\textwidth]{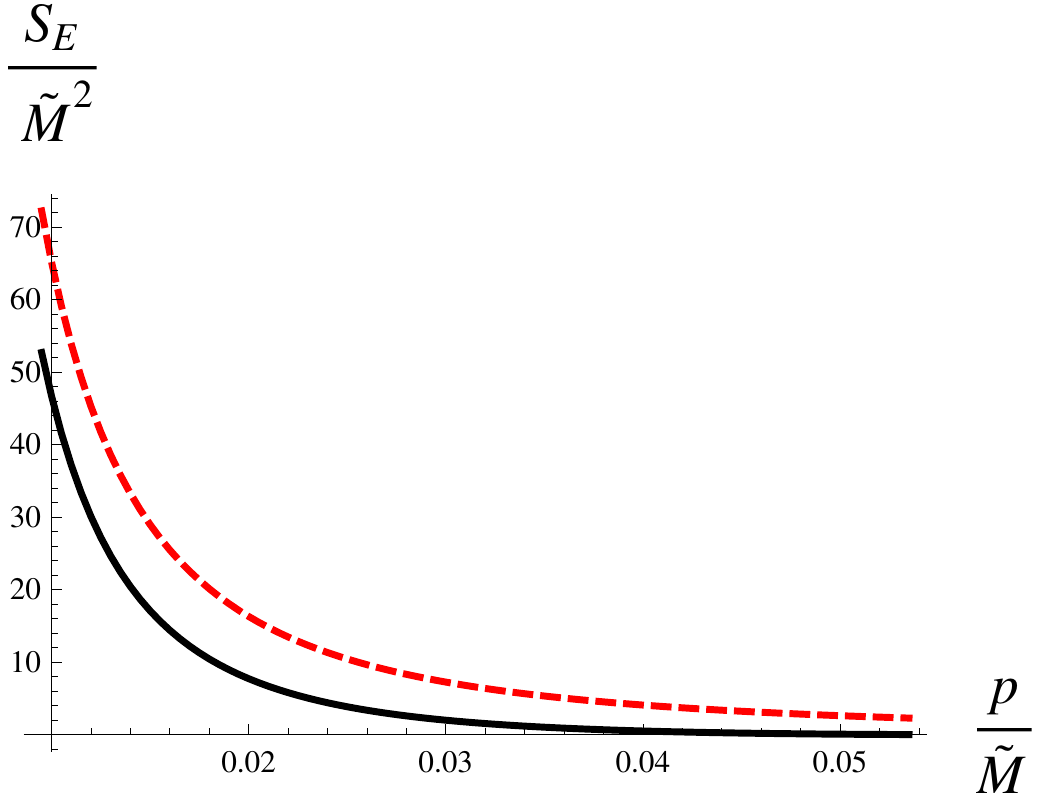}
\caption {The on-shell action for the Euclidean bounce that mediates the decay of the metastable vacuum in the M-theory construction.  The numerical results are shown in black, and the small $p$ approximation \eqref{GotrStarM} is shown in dashed red.
\label {fig:CompareM}}
\end {figure}

When $p$ is not necessarily small, the tunneling coefficient can be computed as in the previous section by finding the $O(3)$-symmetric Euclidean bounce numerically.  We plot the tunneling coefficient at various values of $p/\tilde M$ in figure~\ref{fig:CompareM}.

\section{Discussion}

In this paper we uncovered some new infrared effects in the $(2+1)$-dimensional ${\cal N}=2$ supersymmetric field theory dual to the  $AdS_4\times V_{5,2}$ background of M-theory. We showed that this theory possesses metastable states described by some number of anti-M2 branes placed at the bottom of the
CGLP background \cite{Cvetic:2000db} (this background is a warped product of  $\R^{2, 1}$ and the eight-dimensional Stenzel space \cite{Stenzel} $\sum_{i = 1}^5 {z_i^2} = \epsilon^2$).
We also used semi-classical methods to calculate the decay rates and found them to be highly suppressed for typical parameters.

Our construction is quite analogous to the KPV construction of metastable states \cite{Kachru:2002gs} in the KS background \cite{Klebanov:2000hb}. The gauge theory dual of the KS background is well-understood in terms of a cascade of Seiberg dualities. In particular, an ${\cal N}=1$ supersymmetric $SU(M)\times SU(2M)$ gauge theory provides a good description of the infrared physics \cite{Gubser:2004qj}. Nevertheless, a systematic gauge theory description of the KPV metastable states has not been found. While our M-theory construction of the metastable states is analogous to the KPV construction, 
the dual gauge theory is quite different
\cite{Martelli:2009ga,Jafferis:2009th}: it is $(2+1)$-dimensional and asymptotically conformal in the UV\@.  It would be very interesting to improve our understanding of the effective infrared gauge theory for the CGLP background, and to try describing the metastable states in this context. We have also noted that dimensional reduction to type IIA string theory gives  a warped product of $\R^{1, 1}$ and the eight-dimensional Stenzel space. The curvature of the IIA background is small in the IR\@.  In view of the presence of $\tilde M$ units of R-R flux through the $S^4$ at $\tau=0$, it is tempting to conjecture that the infrared theory is an ${\cal N}=2$ supersymmetric $U(\tilde M)$ gauge theory. We have argued that this gauge theory is not confining because there
are BPS fundamental strings with vanishing effective tension at $\tau=0$. A more detailed understanding of these effects is desirable.

Our M-theory and type IIA arguments for the metastable states were made from the point of view of the M5 and D4-brane world-volume theories, in analogy with the NS5-brane picture used in \cite{Kachru:2002gs}. It should be possible to provide a complementary picture starting with the world-volume gauge theory of $p$ coincident anti-M2 branes. Based on the available results, we expect that these branes would blow up into a fuzzy three-sphere \cite{Basu:2004ed}, but a more detailed investigation of this effect would be interesting.  It would also be useful to find the back-reaction of the $p$ anti-M2 branes on the CGLP background.

\section*{Note added}

After the original version of this paper was submitted, it became clear that the number of M2-branes attached to the M5-brane wrapped over $S^4$ is not exactly $\tilde M$, but rather $\tilde M-1$.  This difference does not significantly affect the  calculations presented in this paper, because these calculations are reliable in the regime $\tilde M\gg 1$ where the difference between $\tilde M$ and $\tilde M-1$ can be neglected.  One way to see that the number of M2-branes is $\tilde M-1$ as opposed to $\tilde M$ is from the number of units of the self-dual $G_4$ flux, which changes from $\tilde M$ on one side of the wrapped M5-brane to $\tilde M-2$ on the other \cite{Hashimoto:2011nn}. As a result, the average flux ``felt'' by the wrapped M5-brane is $\tilde M-1$, and there are $\tilde M-1$ M2-branes attached to it. A consistency check on this result is that the net M2-brane charge at infinity is the same on the two sides of the BPS domain wall of section 4.1:  $\tilde M^2/4 = (\tilde M-2)^2/4 + \tilde M-1$. Similarly, the metastable state with $p$ anti-M2 branes in presence of $\tilde M$ units of $G_4$ flux decays into a supersymmetric vacuum with $\tilde M-1-p$ branes in presence of $\tilde M-2$ units of flux.  We thank Aki Hashimoto and Peter Ouyang for very useful discussions of these issues.

\section*{Acknowledgments}
We thank F.~Benini, A.~Dahlen, A.~Hashimoto, C.~Herzog, T.~Klose, J.~Lin, P.~Ouyang, S.~Sivek, T.~Tesileanu, and H.~Verlinde for useful discussions.  This work was supported in part by the US NSF  under Grant No. PHY-0756966.

\appendix

\section{D4-branes wrapping a 3-sphere at non-zero $\tau$}
\label{BLOWUP}

Recall that the M-theory background is topologically $\R^{2, 1}$ times $\R^4$ fibered over $S^4$, where the radial coordinate of $\R^4$ is $\tau$.  In section~\ref{POTENTIAL} we showed that if we place a small number of anti-M2 branes extended along $(x^0, x^1, x^2)$ at $\tau = \psi = 0$ they tend to blow up into a M5-brane that sits at $\tau = 0$ and at a non-zero value of $\psi$ where it wraps an $S^3 \subset S^4$.

In this section we investigate whether anti-M2 branes at $\tau = \psi = 0$ filling the $(x^0, x^1, x^2)$ directions could blow up into an M5-brane which sits at $\psi = 0$ and wraps an $S^3$ at a fixed value of $\tau>0$.\footnote{A similar computation was done in \cite{Gubser:2004qj} in a type IIB context.}  We again find it convenient to answer this question in the dimensionally reduced type IIA background of section~\ref{IIAREDUCTION} by computing the potential as a function of $\tau$ for a D4-brane with $p$ units of anti-fundamental string charge extended along $(x^0, x^1)$ and wrapping the $S^3 \subset \R^4$ at fixed $\tau$. The fundamental string charge comes from a world-volume electric field ${\cal F}_{01}$ on the D4-brane, which we write as
\es{F01NormalizationTilde}{
{\cal F}_{01} = {1 \over 2 \pi \alpha'} {{\cal E} \over H} \,.
}

In the notation introduced in section~\ref{REVIEW}, the $S^3$ wrapped by this D4-brane has volume form $\vol_{\tilde S^3} = \tilde \sigma_1 \wedge \tilde \sigma_2 \wedge \tilde \sigma_3$ \cite{Martelli:2009ga}.  The relevant component of the form $\alpha$ is
\es{alphaRelevant}{
{3 \over 2} {b^3 c \over \epsilon^3 \cosh^4 {\tau\over 2}} d\tau \wedge
 \tilde \sigma_1 \wedge \tilde \sigma_2 \wedge \tilde \sigma_3
 = {27 \over 8} {\sinh^3 {\tau\over 2} \over \cosh^4 {\tau\over 2}}
 d\tau \wedge
 \tilde \sigma_1 \wedge \tilde \sigma_2 \wedge \tilde \sigma_3 \,,
}
so integrating $C_3$ over this $S^3$ we obtain
\es{C3C5Relevant}{
\int_{S^3} C_3 = {9 m \Vol(S^3) \over 4 \cosh^3 {\tau\over 2}}
 \left[1 - 3 \cosh^2 {\tau\over 2} + 2 \cosh^3 {\tau\over 2} \right] \,.
}
Using \eqref{D4BraneAction} and the observation that $C_5 + C_3 \wedge B_2 = 0$, one finds that the probe D4-brane action for the brane we are interested in is
\es{D4ActionTilde}{
S = \int d^2 x \, {\cal L}_{\cal E} \,,
\qquad
{\cal L}_{\cal E} =  {\mu_4 \epsilon^{9 \over 2} \over m} \Vol(S^3) \left[ -A(\tau)
  \sqrt{1 - (1 + {\cal E})^2} - B(\tau) {\cal E} \right]\,,
}
where
\es{GotAB}{
A(\tau) =  {3^{7 \over 8} (2 + \cosh \tau)^{3\over 8} \sinh^3 {\tau\over 2} \over
2^{3 \over 2} \hat H(\tau)^{1 \over 2}  \cosh^{3 \over 2} {\tau\over 2}}  \,, \qquad
B(\tau) =  {9 \left(1 - 3 \cosh^2 {\tau\over 2} + 2 \cosh^3 {\tau\over 2} \right)
\over 4 \hat H(\tau) \cosh^3 {\tau\over 2}}
 \,,
}
the function $\hat H(\tau)$ being defined in \eqref{GotH}.

As in section~\ref{POTENTIAL}, in order to express the Lagrangian in terms of the fundamental string charge $(-p)$ as opposed to the electric field ${\cal E}$, we need to perform a Legendre transform.  The fundamental string charge is related to the displacement ${\cal D} = {\partial {\cal L}_{\cal E} \over {\partial \cal E}}$ through
\es{DpRelation}{
{\cal D} = {\mu_4 \epsilon^{9 \over 2} \over \hat H m} {9 (-p) \over \tilde M} \,,
}
From the Legendre transformed Lagrangian ${\cal L}_{\cal D} = {\cal L}_{\cal E} - {\cal D} {\cal E}$, one can compute the potential density (potential per unit $x^1$ coordinate length) $V(\tau) = -{\cal L}_{\cal D}$:
\es{LagAgain}{
V(\tau) = \tilde M V_{\rm string}^{(0)}
\left[ \sqrt{{\hat H_0^2 \over 324} A(\tau)^2 +
\left({\hat H_0 \over 18} B(\tau) - {\hat H_0 \over \hat H(\tau)} {p\over 2 \tilde M} \right)^2}
- {\hat H_0 \over 18} B(\tau) + {\hat H_0 \over \hat H(\tau)} {p\over 2 \tilde M}  \right] \,.
}
We expressed $V(\tau)$ as a multiple of the potential $V_{\rm string}^{(0)}$ for an anti-fundamental string placed at $\tau = \psi = 0$ that was computed in eq.~\eqref{Vstring}.  It is not hard to see that since at $\tau = 0$ the functions $A$ and $B$ vanish while $\hat H(0) = \hat H_0$ by definition, we have $V(0) = 0$ if $p\leq 0$ and $V(0) = p V_{\rm string}^{(0)}$ if $p>0$, confirming that our D4-brane contains $(-p)$ units of fundamental string charge spread over its world-volume.

Plotting $V(\tau)$ for various values of $p / \tilde M$ one can check that this function has only one minimum at $\tau = 0$ regardless of $p/\tilde M$.

\bibliographystyle{ssg}
\bibliography{metastable}

\end{document}